\documentclass[aps,prb,preprint,showpacs,byrevtex]{revtex4}
\usepackage{times,mathptm}
\usepackage[dvips]{graphicx,color}

\begin{document}
\title{Antiferromagnetic Slater Insulator Phase of Na$_2$IrO$_3$}
\author{Hyun-Jung Kim}
\author{Jun-Ho Lee}
\author{Jun-Hyung Cho}\email{chojh@hanyang.ac.kr}\thanks{Corresponding author}
\affiliation{Department of Physics and Research Institute for Natural Sciences, Hanyang University,
17 Haengdang-Dong, Seongdong-Ku, Seoul 133-791, Korea}
\date{\today}

\begin{abstract}
Using a hybrid density-functional theory (DFT) calculation including spin-orbit coupling (SOC), we predict that the zigzag antiferromagnetic (AFM) ground state of the honeycomb layered compound Na$_2$IrO$_3$ opens the observed insulating gap through a long-range magnetic order. We show that the effect of SOC and the correction of self-interaction error inherent in previous local or semilocal DFT calculations play crucial roles in predicting the band gap formation in Na$_2$IrO$_3$. It is revealed that the itinerant AFM order with a strong suppression of the Ir magnetic moment is attributed to a considerable hybridization of the Ir 5$d$ orbitals with the O 2$p$ orbitals. Thus, our results suggest that the insulating phase of Na$_2$IrO$_3$ can be represented as a Slater insulator driven by itinerant magnetism.
\end{abstract}


\maketitle


Recently, there have been intensive studies concerning the interplay between the spin-orbit coupling (SOC), the on-site Coulomb repulsion ($U$), and the bandwidth ($W$) in 5$d$ transition metal oxides (TMO)~\cite{okada,pesin,jackeli,shitade,okamoto,machida,gret,singh1,singh2,ohgushi,bjkim1,bjkim2,comin,chaloupka}.  Compared to 3$d$ orbitals in 3$d$ TMO, 5$d$ orbitals in 5$d$ TMO are spatially more extended to yield a relatively smaller (larger) value of $U$ ($W$). On the other hand, the strength of the SOC in 5$d$ TMO is enhanced due to the higher atomic numbers of transition metals. The resulting energy scales arising from SOC, $U$, and $W$ in 5$d$ TMO may become comparable to each other, leading to a broad spectrum of exotic quantum phases such as topological insulator~\cite{shitade,pesin}, spin liquid~\cite{machida,okamoto,chaloupka}, and Mott insulator~\cite{bjkim1,bjkim2,ohgushi,jackeli,singh2,singh1,gret,comin}. In particular, the nature of the insulating phase in iridium (Ir) oxides has been under intense debate whether it is a Mott-type insulator~\cite{bjkim1,bjkim2} or a Slater-type insulator~\cite{arita,li}. Here, the gap formation in the Mott-type insulator is driven by electron correlation while that in the Slater-type insulator is associated with magnetic ordering.

\begin{figure}[ht]
\centering{ \includegraphics[width=7.7cm]{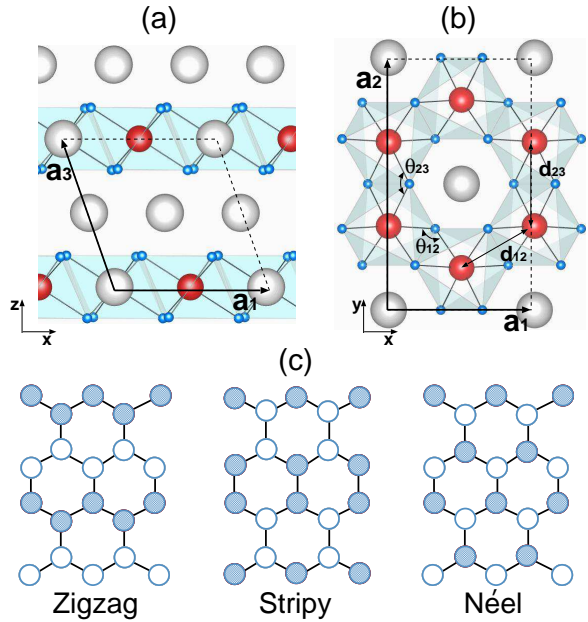} }
\caption{Crystal structure of Na$_2$IrO$_3$: (a) projection on the $xz$ plane and (b) projection on the $xy$ plane. {\bf a}$_1$, {\bf a}$_2$, and {\bf a}$_3$ denote unit vectors of the unit cell. The large, medium, and small circles represent Na, Ir, and O atoms, respectively. Three different AFM structures with the zigzag, stripy, and N$\acute{\rm e}$el spin orders are schematically shown in (c), where the solid and open circles indicate up and down spins.}
\end{figure}

As a prototypical example of 5$d$ TMO, we here focus on Na$_2$IrO$_3$, where Ir atoms form a honeycomb lattice and each Ir atom is surrounded by an octahedron of six O atoms [see Fig. 1(a) and 1(b)]. It was experimentally observed that Na$_2$IrO$_3$ has the antiferromagnetic (AFM) insulating ground state with a zigzag spin alignment [Fig. 1(c)] below the N$\acute{\rm e}$el temperature $T_N$ ${\simeq}$ 15 K~\cite{liu,ye,choi,comin} but the insulating gap is preserved even at room temperature~\cite{comin}.  This separation between the insulating behavior and the onset of AFM ordering may imply that Na$_2$IrO$_3$ can be regarded as a Mott insulator driven by electron correlations~\cite{gret,comin,singh2,liu}. To address the nature of the insulating phase in Na$_2$IrO$_3$, it was suggested that the Ir$^{4+}$ $t_{2g}$ states locating around the Fermi energy would be treated in terms of relativistic atomic orbitals with the effective angular momentum $j_{\rm eff}$ = 1/2 and $j_{\rm eff}$ = 3/2~\cite{shitade,comin,gret,singh2}. Here, the upper $j_{\rm eff}$ = 1/2 band decoupled from the lower $j_{\rm eff}$ = 3/2 bands was presumed to be half-filled, and therefore Ir atoms can have localized magnetic moments corresponding to an effective spin one-half Ir$^{4+}$ ion~\cite{jackeli,chaloupka}. This $j_{\rm eff}$ scenario caused by strong SOC leads not only to a novel magnetic structure suggested from the Kitaev-Heisenberg model~\cite{chaloupka,singh1,chaloupka2} but also a spin-orbit Mott insulator where the half-filled $j_{\rm eff}$ = 1/2 band splits into two Hubbard bands by taking into account on-site Hubbard $U$~\cite{comin,gret,singh2}. However, contrasting with such highly localized $j_{\rm eff}$ = 1/2 orbitals at Ir atoms, a recent $ab$ $initio$ density functional theory (DFT) calculation together with the tight-binding model analysis~\cite{mazin} showed that the $t_{2g}$ bands can be described by quasimolecular orbitals (QMOs) which are fully delocalized over six Ir atoms forming a honeycomb lattice. Despite this rather itinerant character of QMOs, the explicit treatment of $U$ was required to obtain the observed insulating gap of ${\sim}$0.34 eV~\cite{mazin,gret,comin}. It is thus likely that the effect of electron correlations plays an indispensable role in describing the insulating phase of Na$_2$IrO$_3$, thereby being represented as a Mott insulator~\cite{comin,gret,singh2}.

By contrast, we here propose a different mechanism for the observed insulating phase of Na$_2$IrO$_3$ based on a long-range magnetic order. This magnetically driven insulating phase through an itinerant single-particle approach can be represented as a Slater insulator~\cite{slater}. It is noticeable that a rather delocalized character of the Ir $t_{2g}$ states, as described by QMOs~\cite{mazin}, may be associated with the self-interaction error (SIE) inherent to the conventional DFT calculations with the local density approximation (LDA)~\cite{lda} or the generalized gradient approximation (GGA)~\cite{pbe}. Note that the SIE causes the electron density to artificially spread out because delocalization reduces the spurious self-repulsion of electron~\cite{sie,mori}. This so-called delocalization error tends to give an inaccurate estimation of the ionization energy and the electron affinity, resulting in the underestimation of band gap~\cite{mori}. In this regard, previous LDA and GGA calculations~\cite{gret,comin,mazin,liu} may not adequately describe the insulating phase of Na$_2$IrO$_3$. Therefore, it is very challenging to examine how the electronic properties of Na$_2$IrO$_3$ can be changed by the correction of SIE with an exchange-correlation functional beyond the LDA or GGA.

In this paper, we present a new theoretical study of Na$_2$IrO$_3$ based on the hybrid DFT scheme including SOC. We find that the effect of SOC and the correction of SIE with the screened hybrid exchange-correlation functional of Heyd-Scuseria-Ernzerhof (HSE)~\cite{hse1,hse2} opens the observed~\cite{comin} insulating gap of ${\sim}$0.34 eV for the zigzag AFM ground state. Compared with the effective spin $S$ = 1/2 moments ($M$ = 1 ${\mu}_B$) within the Kitaev-Heisenberg model~\cite{chaloupka,singh1}, the calculated magnetic moment per Ir atom is much reduced to be 0.37 ${\mu}_B$, close to that (${\sim}$0.22 ${\mu}_B$) measured by a combined neutron and x-ray diffraction (XRD) experiment~\cite{ye}. Such an itinerant character of magnetism is revealed to be due to a considerable hybridization of the Ir 5$d$ orbitals with the O 2$p$ orbitals.

\section*{\large Calculation details}
Our hybrid DFT calculations including SOC were performed using the Vienna $ab$ $initio$ simulation package (VASP) with the projector augmented wave (PAW) method~\cite{vasp1,vasp2}. For the exchange-correlation energy, we employed the HSE functional~\cite{hse1,hse2}, which is given by
\begin{eqnarray}
E_{XC}^{\rm HSE}  =&&{\alpha}E_{X}^{\rm HF,SR}(\omega) + (1-{\alpha})E_{X}^{\rm PBE,SR}(\omega) \nonumber\\
             &&+ E_{X}^{\rm PBE,LR}(\omega) + E_{C}^{\rm PBE}.
\end{eqnarray}
Here, the mixing factor ${\alpha}$ controls the amount of exact Fock exchange energy and the screening parameter ${\omega}$ (= 0.20 {\AA}$^{-1}$) defines the separation of short range (SR) and long range (LR) for the exchange energy. Note that the HSE functional with ${\alpha}$ = 0 becomes identical to the Perdew-Burke-Ernzerhof (PBE)~\cite{pbe} functional. We used the experimental lattice constants $a$ = 5.427, $b$ = 9.395, and $c$ = 5.614 {\AA} with the $C2/m$ monoclinic crystal structure, obtained by an XRD study~\cite{choi}. The ${\bf k}$-space integration was done with the 6${\times}$3${\times}$6 uniform meshes in the Brillouin zone. All atoms were allowed to relax along the calculated forces until all the residual force components were less than 0.02 eV/{\AA}.

\section*{\large Results and discussion}

\begin{figure}[hb]
\centering{ \includegraphics[width=7.7cm]{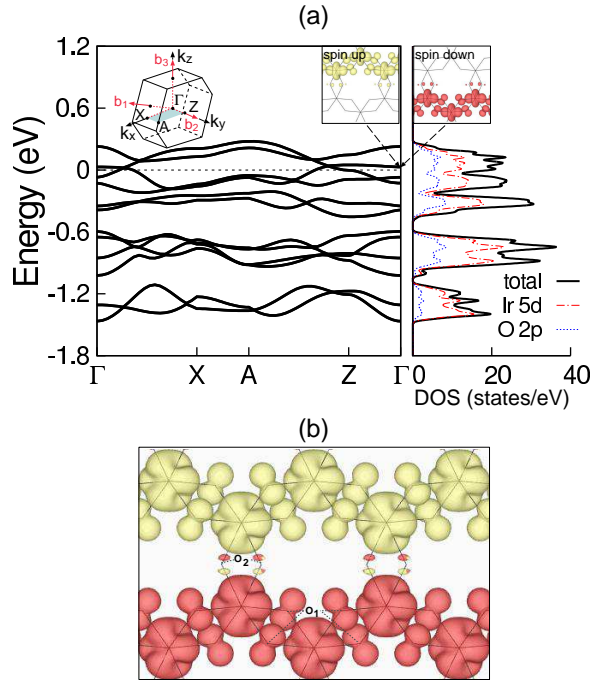} }
\caption{(a) Band structure and DOS of the zigzag AFM structure, obtained using the PBE functional. The band dispersions are plotted along the symmetry lines shown in the Brillouin zone of the unit cell (see the inset). The total DOS, Ir 5$d$ partial DOS, and O 2$p$ partial DOS are displayed with solid, dot-dashed, and dotted lines, respectively. The energy zero represents the Fermi level. The charge characters of the spin-up and spin-down $t_{2g}$ state near the Fermi level are shown with an isosurface of 0.004 $e$/{\AA}$^3$. In (b), the spin-up (spin-down) density is displayed in bright (dark) color with an isosurface of 0.004. ($-$0.004) $e$/{\AA}$^3$, and the two different species of O atoms are denoted as O$_1$ (for O atoms on the same zigzag chain side) and O$_2$ (for O atoms between two zigzag chains).}
\end{figure}

We begin to study the experimentally observed~\cite{liu,ye,choi} zigzag AFM ground state of Na$_2$IrO$_3$ using the PBE calculation. The optimized structural parameters such as the Ir-Ir bond lengths [$d_{12}$ and $d_{23}$ in Fig. 1(b)] and the Ir-O-Ir bond angles [${\theta}_{12}$ and ${\theta}_{23}$ in Fig. 1(b)] are given in Table I. The calculated values of ${\theta}_{12}$ = 99.84$^\circ$ and ${\theta}_{23}$ = 98.51$^\circ$ (greater than the ideal 90$^\circ$ Ir-O-Ir bond angle) show a sizable trigonal distortion of the IrO$_6$ octahedra, consistent with an XRD analysis~\cite{choi}. Figure 2(a) shows the calculated band structure and density of states (DOS) of the zigzag AFM structure, which exhibit the presence of partially occupied Ir $t_{2g}$ states at the Fermi level ($E_{F}$), indicating a metallic feature. Note that there are twelve $t_{2g}$ bands which originate from four different Ir atoms within the unit cell of the zigzag AFM structure. It is noticeable that, for the $t_{2g}$ states locating near $E_F$, the partial DOS projected onto the O 2$p$ orbitals amounts to ${\sim}$50\% of that projected onto the Ir 5$d$ orbitals [see Fig. 2(a)], indicating a considerable hybridization between the two orbitals. Indeed, the spin characters of the $t_{2g}$ state at $E_F$, as shown in Fig. 2(a), reveal the electron delocalization over IrO$_6$ octahedra on each zigzag chain side. These aspects of the $t_{2g}$ states lead not only to a smaller magnetic moment of ${\sim}$0.53 ${\mu}_B$ for an Ir atom compared to the effective spin $S$ = 1/2 moment assumed in the Kitaev-Heisenberg model~\cite{chaloupka,singh1} but also a slightly induced magnetic moment of ${\sim}$0.10 ${\mu}_B$ for the O$_1$ atom [see Fig. 2(b) and Table II]. It is noteworthy that the PBE calculation may involve the over-delocalization of the $t_{2g}$ states due to the SIE, therefore incorrectly predicting the zigzag AFM structure to be metallic rather than insulating.

In order to correct the SIE, we use the HSE functional to optimize the zigzag AFM structure. We find that the structural parameters slightly depend on the magnitude of ${\alpha}$ in the HSE functional (see Table I): i.e., $d_{12}$ and $d_{23}$ change little by less than 0.01 {\AA} in the range of 0 $<$ ${\alpha}$ ${\leq}$ 0.08, but, as ${\alpha}$ increases to 0.1, $d_{12}$ ($d_{23}$) decreases (increases) by ${\sim}$0.02 (0.04) {\AA}. We also find a drastic variation of the band structure and DOS as a function of ${\alpha}$. Especially, the DOS at $E_F$ is found to decrease as ${\alpha}$ increases up to 0.08, giving rise to the creation of a pseudogap. For ${\alpha}$ $>$ 0.08, the pseudogap is turned into an insulating gap whose magnitude increases with increasing ${\alpha}$ (see Fig. 3). Figure 4(a) shows the band structure and DOS of the zigzag AFM structure, obtained using the HSE calculation with ${\alpha}$ = 0.05. It is seen that, compared with the PBE result [Fig. 2(a)], (i) the DOS for the occupied $t_{2g}$ states is shifted to a lower energy and (ii) the DOS at $E_F$ is much reduced. We note that, as ${\alpha}$ increases, the Ir magnetic moment increases compared to that obtained using the PBE calculation (see Table II), reflecting that the HSE functional corrects the over-delocalization of Ir 5$d$ electrons due to the SIE of the PBE functional.

\begin{figure}[ht]
\centering{ \includegraphics[width=7.7cm]{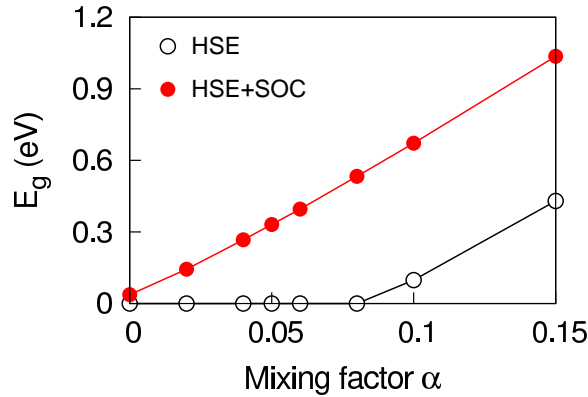} }
\caption{Calculated band gap of the zigzag AFM structure as function of the mixing factor ${\alpha}$ in the HSE functional.}
\end{figure}

According to the $j_{\rm eff}$ scenario, the SOC splits the $t_{2g}$ states into the $j_{\rm eff}$ = 1/2 and $j_{\rm eff}$ = 3/2 states~\cite{shitade,chaloupka,jackeli}, and the experimental 0.34 eV-gap was opened by taking into account an on-site interaction $U$ of 1$-$3 eV, leading to a conclusion that the insulating phase of Na$_2$IrO$_3$ can be represented as a spin-orbit Mott insulator~\cite{comin,gret,singh2}. In this study, we examine the effect of SOC on the electronic structure of the zigzag AFM structure using the HSE+SOC calculation. Here, we employ the optimized HSE structure because the effect of SOC changes little the Ir-Ir bond lengths and the Ir-O-Ir bond angles by less that 0.01 {\AA} and 1$^\circ$, respectively. As shown in Fig. 3, the inclusion of SOC opens the band gap $E_g$, which monotonically increases with increasing ${\alpha}$. For ${\alpha}$ = 0 (equivalent to the PBE+SOC calculation), we obtain $E_g$ = 0.05 eV, in good agreement with a previous PBE+SOC calculation~\cite{sohn}. On the other hand, as ${\alpha}$ increases to 0.05, $E_g$ becomes 0.33 eV, close to the experimental value of ${\sim}$0.34 eV~\cite{comin}. The HSE+SOC band structure and DOS computed with ${\alpha}$ = 0.05 are displayed in Fig. 4(b). It is seen that (i) the $t_{2g}$ bands just below and above $E_F$ become almost dispersionless, thereby possibly correcting the over-delocalization of the $t_{2g}$ states, and (ii) there are six separated energy regions for the $t_{2g}$ bands, consistent with the experimental observation of the five $d$-$d$ interband transitions in optical conductivity~\cite{sohn}. Since the HSE+SOC calculation with ${\alpha}$ = 0.05 adequately predicts the insulating electronic structure of the zigzag AFM ground state, we can say that the insulating phase of Na$_2$IrO$_3$ can be represented as a spin-orbit Slater insulator through itinerant magnetism. We note that the optimal HSE+SOC value of ${\alpha}$ = 0.05 reproducing the experimental 0.34 eV-gap~\cite{comin} is smaller than that (${\alpha}$ ${\approx}$ 0.15) obtained from the HSE calculation (see Fig. 3). This indicates that the HSE calculation needs larger ${\alpha}$ to cure over-delocalization of the $t_{2g}$ states due to relatively larger SIE.

\begin{figure}[ht]
\centering{ \includegraphics[width=7.7cm]{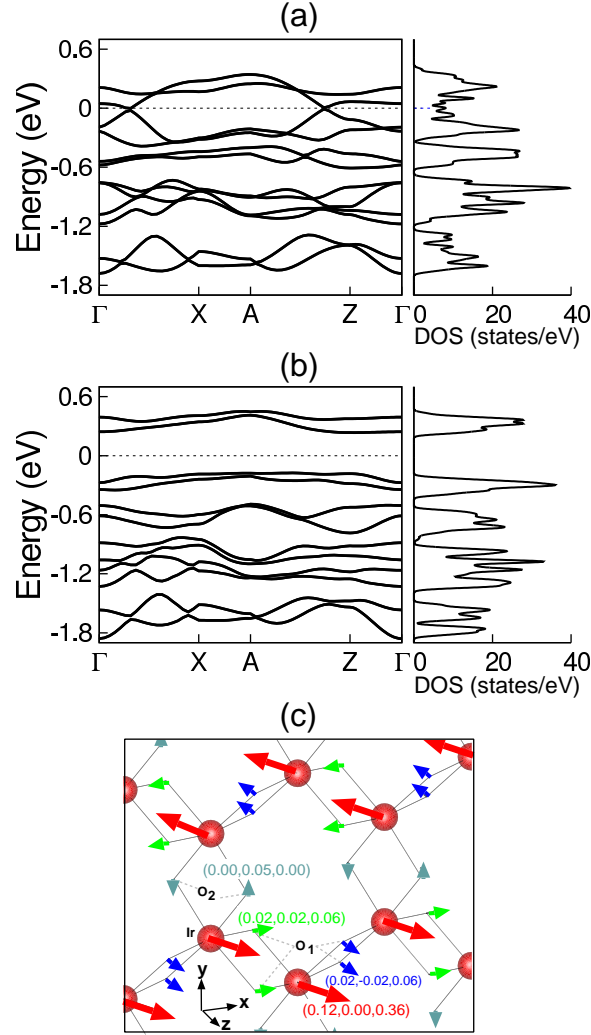} }
\caption{Band structure and DOS of the zigzag AFM structure, obtained using (a) the HSE and (b) HSE+SOC calculations with ${\alpha}$ = 0.05. The magnetic moment ($M_x$, $M_y$, $M_z$), obtained using the HSE+SOC calculation with ${\alpha}$ = 0.05, is drawn in (c). Here, $M_i$ is calculated by integrating the corresponding component of magnetic moment inside the PAW sphere with a radius of 1.4 (0.8) {\AA} for Ir (O). In (c), the circles represent Ir atoms.}
\end{figure}

Figure 4(c) shows the magnetic moment ($M_x$, $M_y$, $M_z$) for each atom of the zigzag AFM structure, obtained using the HSE+SOC calculation with ${\alpha}$ = 0.05. We find that Ir, O$_1$, and O$_2$ atoms have the three components of magnetic moment (${\pm}$0.12, 0.00, ${\pm}$0.36), (${\pm}$0.02, ${\pm}$0.02, ${\pm}$0.06) or (${\pm}$0.02, ${\mp}$0.02, ${\pm}$0.06), and (0.00, ${\pm}$0.05, 0.00) in units of ${\mu}_B$, respectively. It is notable that, when compared with the HSE calculation with the same ${\alpha}$ = 0.05, the inclusion of SOC reduces the magnitude $M$ = $\sqrt {{M_x}^2+{M_y}^2+{M_z}^2}$ for Ir (O$_1$) atom by ${\sim}$34(46) $\%$ but creates $M_y$ = ${\pm}$0.05 ${\mu}_B$ for O$_2$ atom [see Table II and Fig. 4(c)]. Thus, the HSE+SOC value of $M$ = 0.37 ${\mu}_B$ for Ir atom is closer to that (${\sim}$0.22 ${\mu}_B$) measured by a combined neutron and XRD experiment~\cite{ye}. These theoretical and experimental evidences for a strong suppression of the Ir magnetic moment support a Slater-type insulator via itinerant magnetism rather than localized magnetism proposed by the Kitaev-Heisenberg model with effective $S$ = 1/2 pseudospins~\cite{chaloupka,singh1,chaloupka2}. We note that the HSE calculations for half-filled systems~\cite{giovan,lee} have highly debated about the distinction between a Mott-type insulator and a Slater-type insulator. For instance, the HSE calculation for the Cs$_3$C$_{60}$ crystal predicted~\cite{giovan} not only a highly localized half-filled state but also a magnetic moment of ${\sim}$1 ${\mu}_B$ corresponding to a $S$ = 1/2 state, thereby being characterized as a Mott-type insulator. On the other hand, the HSE calculation for the Sn/Ge(111) surface predicted~\cite{lee} a rather delocalized half-filled state with a much reduced magnetic moment of ${\sim}$0.2 ${\mu}_B$, suggesting a  Slater-type insulator.

It was experimentally observed that the paramagnetic insulating phase~\cite{singh1,singh2} of Na$_2$IrO$_3$ exists between $T_N$ ${\simeq}$ 15 K and room temperature~\cite{comin}. However, the precise nature of the paramagnetic insulating phase above $T_N$ is still open to question. The experimental evidence that $T_N$ is much smaller than the Weiss temperature~\cite{singh1,singh2} (${\approx}$ $-$120 K) may indicate a frustration of the underlying AFM interactions. We note that the stripy and N$\acute{\rm e}$el AFM structures [see Fig. 1(c)] can be feasible to model the paramagnetic phase because these AFM structures and the paramagnetic phase have the common features such as similar Ir magnetic moments and zero net magnetic moments in their unit cells. For the stripy and N$\acute{\rm e}$el AFM structures, we perform the HSE+SOC calculations with ${\alpha}$ = 0.05. The stripy (N$\acute{\rm e}$el) AFM structure is found to be not only less stable than the zigzag AFM structure by 4.0 (12.5) meV per Ir atom, but also insulating with $E_{\rm g}$ = 0.18 (0.30) eV. However, we were not able to obtain the nonmagnetic structure which was always converged to the N$\acute{\rm e}$el AFM structure. Since the energy differences among the zigzag, stripy, and N$\acute{\rm e}$el AFM structures are very small, not only is $T_N$ much lower compared to the cases of other iridates but quantum fluctuations of such AFM structures are also likely present above $T_N$, leading to an effectively paramagnetic phase. Further experimental and theoretical studies for the paramagnetic insulating phase are demanded to resolve its nature.

\section*{\large Conclusion}
We performed the HSE+SOC calculation to investigate the zigzag AFM ground state of Na$_2$IrO$_3$. We found that the observed insulating gap of ${\sim}$0.34 eV~\cite{comin} is well predicted by not only taking into account the SOC but also correcting the SIE inherent in the LDA or the GGA. Thus, our results indicate a significant Slater-type character of gap formation through itinerant magnetism. As a matter of fact, we predicted a small itinerant magnetic moment of 0.37 ${\mu}_B$ per Ir atom, contrasting with a fully localized magnetic moment of 1 ${\mu}_B$ per Ir atom presumed within the Kitaev-Heisenberg model~\cite{chaloupka,singh1,chaloupka2}. Similar to the present case of Na$_2$IrO$_3$, we anticipate that the correction of SIE would be of importance to describe the insulating phases of other Ir oxides. We note that other 5$d$ TMO such as Sr$_2$IrO$_4$~\cite{li} and NaOsO$_3$~\cite{calder} were recently reported to display a magnetically driven gap formation, supporting a Slater-type insulator.


\section*{\large Acknowledgement}
This work was supported by National Research Foundation of Korea (NRF) grant funded by the Korean Government (NRF-2011-0015754). The calculations were performed by KISTI supercomputing center through the strategic support program (KSC-2013-C3-006) for the supercomputing application research.

\section*{\large Author contributions}
H.J.K and J.H.L carried out the DFT calculations and data analysis. J.H.C was responsible for the planning and the management of the project. J.H.C wrote the main manuscript text and H.J.K prepared figures 1$-$4. All authors reviewed the manuscript.

\section*{\large Additional information}
\noindent {\bf Competing financial interests:} The authors declare no competing financial interests.

\newpage
\section*{\large Figure legends}
\noindent {\bf Figure 1.} Crystal structure of Na$_2$IrO$_3$: (a) projection on the $xz$ plane and (b) projection on the $xy$ plane. {\bf a}$_1$, {\bf a}$_2$, and {\bf a}$_3$ denote unit vectors of the unit cell. The large, medium, and small circles represent Na, Ir, and O atoms, respectively. Three different AFM structures with the zigzag, stripy, and N$\acute{\rm e}$el spin orders are schematically shown in (c), where the solid and open circles indicate up and down spins.

\vspace{0.4cm}

\noindent {\bf Figure 2.} (a) Band structure and DOS of the zigzag AFM structure, obtained using the PBE functional. The band dispersions are plotted along the symmetry lines shown in the Brillouin zone of the unit cell (see the inset). The total DOS, Ir 5$d$ partial DOS, and O 2$p$ partial DOS are displayed with solid, dot-dashed, and dotted lines, respectively. The energy zero represents the Fermi level. The charge characters of the spin-up and spin-down $t_{2g}$ state near the Fermi level are shown with an isosurface of 0.004 $e$/{\AA}$^3$. In (b), the spin-up (spin-down) density is displayed in bright (dark) color with an isosurface of 0.004. ($-$0.004) $e$/{\AA}$^3$, and the two different species of O atoms are denoted as O$_1$ (for O atoms on the same zigzag chain side) and O$_2$ (for O atoms between two zigzag chains).

\vspace{0.4cm}

\noindent {\bf Figure 3.} Calculated band gap of the zigzag AFM structure as function of the mixing factor ${\alpha}$ in the HSE functional.

\vspace{0.4cm}

\noindent {\bf Figure 4.} Band structure and DOS of the zigzag AFM structure, obtained using (a) the HSE and (b) HSE+SOC calculations with ${\alpha}$ = 0.05. The magnetic moment ($M_x$, $M_y$, $M_z$), obtained using the HSE+SOC calculation with ${\alpha}$ = 0.05, is drawn in (c). Here, $M_i$ is calculated by integrating the corresponding component of magnetic moment inside the PAW sphere with a radius of 1.4 (0.8) {\AA} for Ir (O). In (c), the circles represent Ir atoms.

\newpage
\section*{\large Tables}

\begin{table}[ht]
\caption{Ir-Ir bond lengths [${\theta}_{12}$ and ${\theta}_{23}$ in Fig. 1(b)] and Ir-O-Ir bond angles [${\theta}_{12}$ and ${\theta}_{23}$ in Fig. 1(b)] obtained using the PBE and HSE calculations, in comparison with the experimental data.}
\begin{ruledtabular}
\begin{tabular}{lcccc}
            & $d_{12}$ (\AA)  & $d_{23}$ (\AA) & ${\theta}_{12}$ ($^\circ$) & ${\theta}_{23}$ ($^\circ$) \\ \hline
PBE         &  3.129 & 3.139 & 99.84  & 98.51  \\
HSE$_{{\alpha}=0.05}$  & 3.130 & 3.138 & 100.23  & 98.86  \\
HSE$_{{\alpha}=0.1}$ & 3.109 & 3.180 & 99.51  & 99.73  \\
Experiment (Ref.~\cite{choi})  & 3.130 & 3.138 & 99.45  & 97.97  \\
\end{tabular}
\end{ruledtabular}
\end{table}

\begin{table}[ht]
\caption{Magnitude of the magnetic moments (in units of ${\mu}_B$) of Ir, O$_1$, and O$_2$ atoms obtained using the PBE, HSE, PBE+SOC, and HSE+SOC calculations.}
\begin{ruledtabular}
\begin{tabular}{lccc}
            &  Ir  &  O$_1$  &  O$_2$  \\ \hline
PBE         &  0.53 & 0.10 & 0.00    \\
HSE$_{{\alpha}=0.05}$  & 0.56 & 0.11 & 0.00    \\
HSE$_{{\alpha}=0.1}$  & 0.62 & 0.11 & 0.00    \\
PBE+SOC         &  0.33 & 0.06 & 0.05    \\
HSE+SOC$_{{\alpha}=0.05}$ & 0.37 & 0.06 & 0.05    \\
HSE+SOC$_{{\alpha}=0.1}$ & 0.41 & 0.06 & 0.05    \\
\end{tabular}
\end{ruledtabular}
\end{table}

\end{document}